\documentclass[%
 reprint,
 amsmath,amssymb,
 aps,
floatfix,
]{revtex4-2}
\usepackage{comment}
\usepackage{graphicx}
\usepackage{dcolumn}
\usepackage{bm}
\usepackage{aas_macros} 
\usepackage{color} 
\newcommand{\com}[1]{\textcolor{blue}{[{\bf KI}: #1]}}

\newcommand{\hk}[1]{\textcolor{green}{[{\bf HK}: #1]}}
\usepackage{comment}
\usepackage{amsmath}
\usepackage{mathrsfs}
\usepackage{mathcomp}

\def\red#1{\textcolor{red} {#1}}
\def\kakko#1{\left( #1\right)}

\def\KAkko#1{\left[ #1\right]}

\def\Kan#1{\langle #1 \rangle}

\def\bee{\begin{eqnarray}}
\def\ene#1{\label{#1}\end{eqnarray}}

\def\bea{\begin{array}}
\def\ena{\end{array}}
\def\f{\frac}

\def\pol{_{\rm{pol}}}
\def\met{{\rm{mehthod}}}
\def\tot{_{\rm{tot}}}
\def\s{\sigma}
\def \pri{_{\rm{prior}}}
\def\vx{\vec{x}}
\def\hx{\hat x}
\def\vk{\vec{k}}
\def\hn{\hat n}

\def\atm{a_{2m}}
\def\lm{_{lm}}

\def\d{\Delta}

\def\ri{R_{\rm{ini}}}
\def\mc#1{{\mathcal{#1}}}

\def\clu{_{\rm{cluster}}}
\def\fid{_{\rm{fiducial}}}
\def\sm{\sigma_{\rm{method}}}

\def\uk{\rm{\mu K}}
\def\t{\times}
\def\x{\chi^2}
\def\side{_{\rm{side}}}

\def\e{\tau}

\def \om{\Omega}

\def\l {\Lambda}
\def\inf{\infty}

\def\gcm3{\mathrm{g} / \mathrm{cm}^3}

\newcommand{\cf}{\cfrac}

\def \phii{\phi_{\rm{ini}}}

\def\pa{\partial}

\def\msum{\sum^l_{m=-l}}
\def\lsum{\sum^\inf_{l=0}}

\def\id3k{\int d^3k}
\def\eik{e^{i\vk\cdot\vx}}

\def\vke{(\vk,\e)}

\def\cmb{_{\rm{CMB}}}
\def\tc{\tau_C}

\def\sq#1{\sqrt{#1}}
\def\y22m{{}_2Y_{2m}}

\def\cs2{c_s{}^2}

\def\sec#1{\section{#1}}

\begin{document}

\preprint{APS/123-QED}

\title{Validating dark energy models using polarised Sunyaev-Zel'dovich effect with large-angle CMB temperature and E-mode polarization anisotropies}
\

\author{Hiroto Kondo$^{1}$}
\email{h.kondo@nagoya-u.jp}
\author{Kiyotomo Ichiki$^{1,2}$}
\author{Hiroyuki Tashiro$^{1}$}
\author{Kenji Hasegawa$^{1}$}
\affiliation{%
$^1$Graduate School of Science, Division of Particle and
Astrophysical Science, Nagoya University, Chikusa-Ku, Nagoya, 464-8602, Japan\\
$^2$Kobayashi-Maskawa Institute for the Origin of Particles and the
Universe, Nagoya University, Chikusa-ku, Nagoya, 464-8602, Japan
}%



\date{\today}

\begin{abstract}
The tomography of the polarized Sunyaev-Zeldvich effect due to free electrons of galaxy clusters can be used to constrain the nature of dark energy because CMB quadrupoles at different redshifts as the polarization source are sensitive to the integrated Sachs-Wolfe effect. Here we show that the low multipoles of the temperature and E-mode polarization anisotropies from the all-sky CMB can improve the constraint further through the correlation between them and the CMB quadrupoles viewed from the galaxy clusters. Using a Monte-Carlo simulation, we find that low multipoles of the temperature and E-mode polarization anisotropies potentially improve the constraint on the equation of state of dark energy parameter by $\sim 17$ percent.
\end{abstract}

\maketitle


\section{\label{sec:intro}Introduction}
Dark energy, which is causing the current accelerated expansion of the
universe \cite{1998AJ....116.1009R,1999ApJ...517..565P}, has two main
effects on the temperature anisotropies of the cosmic microwave
background (CMB). One is to change the angular distance to the final
scattering surface of the CMB, and the other is the Integrated
Sachs-Wolfe (ISW) effect, which creates new temperature fluctuations due
to the decay of the gravitational potential of the large-scale structure.  The
ISW effect is a characteristic effect that indicates that the universe
is deviating from the matter-dominated one. However, because the
temperature fluctuations produced by this effect are smaller than those
produced in the early universe in the standard cosmological model
(so-called the SW effect), they are masked by the dispersion of the
fluctuations, making it difficult to obtain a statistically significant
enough signal to approach the nature of dark energy.  Therefore, the CMB
constraint on dark energy-related parameters is weak because the ISW
effect suffers from sizable cosmic variance errors in the CMB
temperature anisotropy spectrum on large scales
\cite{2013ApJS..208...19H,2020A&A...641A...6P}.

The Kamionkowski and Loeb method \cite{1997PhRvD..56.4511K} is an
effective way to detect the ISW effect without this cosmic
variance. This method uses the fact that the polarization angle of CMB
photons scattered by free electrons in a galaxy cluster is determined by
the quadrupole temperature fluctuations of the CMB as seen from that
cluster \cite{2000PhRvD..62l3004S} and allow us to reconstruct the three-dimensional
density fluctuations of the universe on large scales
\cite{2004PhRvD..70f3504P,2006PhRvD..73l3517B,2007PhRvD..75j1302A,2016MNRAS.460L.104L}. While
avoiding cosmic variance by fixing the realization of the initial
density fluctuations, the direct detection of the ISW effect is possible
by tomographic use of clusters of galaxies at various redshifts
\cite{2003PhRvD..67f3505C,2004PhRvD..69b7301C,2005PhRvL..95j1302S}. Our previous study using simple Monte Carlo simulations has shown that it is possible to constrain the dark energy equation of state parameters
more accurately through the ISW effect than conventional methods based on the power spectra 
\cite{2022PhRvD.105f3507I}.
The method can also be useful for the studies of
the power asymmetry of CMB polarization and density field
\cite{2018JCAP...04..034D}, cosmic birefringence \cite{2022PhRvD.106h3518L} and the reionization optical depth \cite{2018PhRvD..97j3505M}.

In our previous study \cite{2022PhRvD.105f3507I}, we used the quadrupole anisotropies of the CMB as a diagnostic of the ISW effect. Specifically, we compared the quadrupole anisotropy of our CMB estimated from the three-dimensional density fluctuations on large scales reconstructed by the KL method, with the actual quadrupole anisotropy that
can be directly observed by the all-sky CMB experiments such as Planck. 
In fact, it has
been shown that the three-dimensionally
reconstructed density fluctuations on large scales should be correlated
not only with the quadrupoles but also with higher temperature multipoles and E-mode polarization fluctuations on large angular scales \cite{2017PhRvD..96l3509L}. Therefore, this paper aims to clarify to what
extent the addition of these fluctuations as diagnostics improves the results obtained
in previous studies.

In the next section, we review our method developed in \cite{2022PhRvD.105f3507I}, and extend it by adding information on the temperature and E-mode polarization anisotropies on large angular scales. Section III presents our result of the future constraint on the dark energy equation of state parameters based on Monte-Carlo simulations. In Section IV, we discuss and summarize this study.
\section{Methodology}

\subsection{CMB polarization from galaxy clusters}\label{polarization}

First, we consider the CMB polarization produced in galaxy clusters.
The polarization is created by Thomson scattering of the free electrons in galaxy clusters with the quadrupole component of the CMB anisotropy.
Therefore, if a galaxy cluster is at the position $\vx$ 
in the comoving coordinate, we can observe the polarization from the galaxy cluster, which is produced by the Thomson scattering at the conformal time $\tau_x = \tau_0 -|\vx|$ with the present conformal $\tau_0$.

Accordingly, the observed polarization from galaxy clusters at $\vx$
can be calculated with the Stokes parameter, $Q(\vx)$ and $U(\vx)$ 
\bee
Q(\vx)\pm iU(\vx)&=&-\f {\sq 6}{10}\tc T\cmb(\tau_x) \nonumber \\
&&\times  \sum^2_{m=-2}{}_\pm\y22m(\hx)\atm^T (\vx, \tau_x)~,
\ene{p51}
where $\tc$ is the optical depth of the galaxy cluster for Thomson scattering.
In Eq.~\eqref{p51}
$\atm^T (\vx, \tau_x)$ is the quadrupole component of the CMB temperature anisotropy observed at the position
$\vx$ and the conformal time $\tau_x$.

Now we consider the CMB temperature anisotropy
on the position of the comoving coordinate $\vx$ 
at the conformal time $\e_x$.

The CMB temperature in the direction $\hn$ at $\vx$ and $\e_x$ can 
be decomposed into
the isotropic part and anisotropic part,
$T(\vx ,\hn ,\e_x )= T\cmb(\vx ,\e_x ) +\d T(\vx ,\hn ,\e_x )$.
Introducing the CMB anisotropy as
$\d (\vx ,\hn ,\e_x ) \equiv  {\d T(\vx ,\hn ,\e_x )}/{T\cmb}$
and we expand the CMB anisotropy with sperical harmonic functions $Y_{lm}(\hn )$,
\bee
\d (\vx ,\hn ,\e_x ) =\lsum\msum a^T_{lm}(\vx ,\e_x)Y_{lm}(\hn )~,
\ene{ylm-decompose}
where $a^T_{lm}(\vx ,\e)$ is the coefficient of the spherical harmonic expansion
and the coefficinet with $\ell =2$ is the quadrupole component $\atm^T (\vx, \tau_x)$. 

On the other hand, since the CMB anisotropy is the function of $\vx$ and $\hn$,
we can decompose it by the plane wave function and the spherical harmonics,
\bee
\d (\vx,\hn,\e)&=&4\pi\id3k \eik \lsum (-i)^l\d^T_l\vke \nonumber \\
&&\times \msum Y\lm^*({\hat k})Y\lm(\hn)~.
\ene{p14}
Therefore, the coefficient of the spherical harmonic expansion in Eq.~\eqref{ylm-decompose}
can be written as 
\bee
a^T\lm(\vx,\e_x)
=(-i)^l4\pi\id3k\eik\d_l^T({\vk, \e_x}) Y\lm^*({\hat k})~.
\ene{p15}

In our case, the cosmological linear perturbation theory is applicable to
calculate $\d^T_l\vke$ in Eq.~\eqref{p14}.
According to the cosmological linear perturbation theory,
$\d^T_l\vke$ can be obtained as 
\bee
\d^T_l\vke =\d^T_l(k,\e)\phii (\vk)~,
\ene{p17}
where $\phii (\vk)$ is the Fourier component of the initial curvature perturbations,
and $\d^T_l(k,\e)$ is the liner transfer function 
which depends on the cosmological models and is obtained from the cosmological linear perturbation theory. We calculate $\d^T_l(k,\e)$ using a publicly available code CAMB  \cite{Lewis:1999bs}.

\subsection{Monte-Carlo simulation}

Our aim of this paper is to study how much the KL methods with
the future CMB temperature and polarization measurement improve the constraint
on the nature of dark energy.
For this purpose, we demonstrate the KL methods by conducting the Monte Carlo simulation.

In our simulation, to realize the CMB anisotropy at comoving position $\vx$ 
we use transfer functions generated by the publicly available code CAMB. 
Throughout this paper, we set $\l$-CDM model with $\om_bh^2=0.0226$,  $\om_ch^2=0.112$, $\om_\nu h^2=0.00064$, $h=0.7$, as the reference cosmological models.

The first step of the simulation is to generate the initial fluctuation field $\phii (k_i)$. 
Our initial fluctuation field is given as a Gaussian random field with 
the power spectrum
\bee
\mc P(k)=\f{k^3}{2\pi^2}P(k)=A_s\kakko{\f k {k_*}}^{n_s-1}~,
\ene{i5}
where we set the parameters $A_s=2.1 \times 10^{-9}$, $n_s=0.96$ and $k_*=0.05$. 

In our methods, it is useful to employ the polar coordinate 
in Fourier $k$-space.
To sample the Fourier mode, 
we divide the angular directions in Fourier space by Healpix \cite{2005ApJ...622..759G} with $N_{\rm{side}}=8$.
This means that the whole sky is divided into 768 sections.
For the radial mode,
we sample 60 wave number modes uniformly in logathmical space
with a range from $k=10^{-5}$ to $10^{-1}$.
Thus, the overall independent Fourier mode~$n_k$ for this simulation is 46080.

Second, we simulate the polarization produced in clusters 
with the generated initial fluctuations $\phii(k_i)$.
In this process, we use the transfer function $\d(k,\e_x)$ with the fiducial equation of state of dark energy parameter $w=-1$.
In our simulation, we set the number of galaxy clusters to $N\clu =6000$.
We distribute them randomly in the angular direction and uniformly in redshift ranging from $z=0$ to $2$.
We calculate the polarization, $Q\fid(\vx_i)$ and $U\fid(\vx_i)$, produced by each galaxy cluster 
at the position $\vx_i$, following the procedure described in Sec.~\ref{polarization}.
To consider the observational uncertainties, Gaussian noise $\s_{\rm {obs}}/\tau = 10^{-2}\ \uk$ is added to each $Q\fid(\vx_i)$ and $U\fid(\vx_i)$.

Similarly, we simulate the CMB anisotropies directly observed at the origin
with the generated initial fluctuations $\phii(k_i)$.
In both the temperature and the polarization anisotropy~(E-mode), we calculate the 
angular components, $a^T\lm{}\fid$ 
$a^E\lm{}\fid$, in the range from $l=2$ to $9$. 
{Here, as in the case for galaxy cluster polarization, we add Gaussian noise $\s_{\rm {obs}} = 10^{-2}\ \uk$ to $a_{lm}$s as observational uncertainty. }

\begin{figure}[htbp]
\centering
\includegraphics[width=1.\hsize]{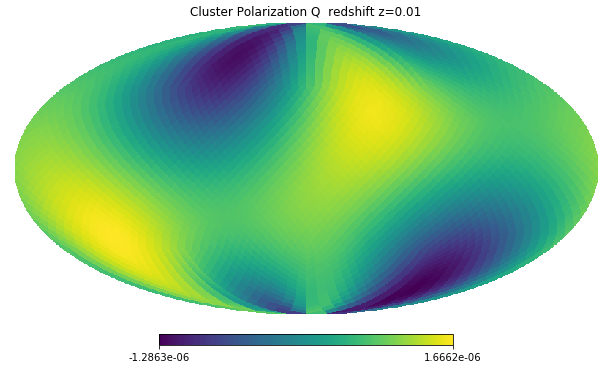}
\caption{Example $Q$ polarization map observed at galaxy clusters at redshift $z=0.01$. Because they are produced by quadrupoles that are nearly identical to the quadrupoles we observe today, they have a quadrupole pattern.}
\label{q001} 
\end{figure}

\begin{figure}[htbp]
\centering
\includegraphics[width=1.\hsize]{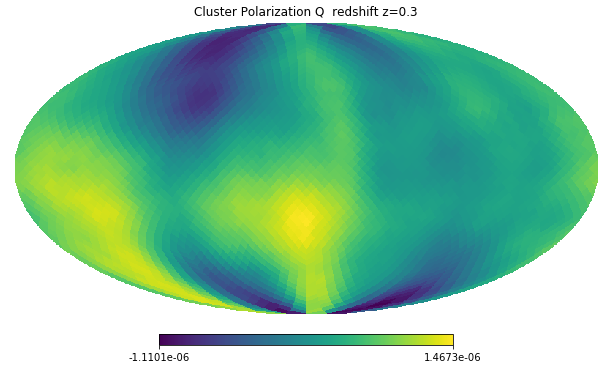}
\caption{Same as Fig.~1, but at redshift $z=0.3$. While features similar to the map at $z=0.01$ remain, smaller patterns develop.}
\label{q030} 
\end{figure}

Fig.\ref{q001} and \ref{q030} show one realization example of the Q maps 
for the polarization produced in galaxy clusters at $z=0.01$ and $0.3$.
The quadrupole of the CMB temperature observed by galaxy clusters at $z=0.01$
is nearly identical to the CMB temperature quadrupole anisotropy at the origin.
Therefore, according to Eq.~\eqref{p51},
the pattern of the Q map on the sky is very similar to that of 
the CMB temperature quadrupole anisotropy at the origin.
On the other hand, at $z=0.3$, the quadrupole pattern observed at each galaxy cluster
is slightly different. Therefore, the generated Q map has small-scale pattern due to the difference,
although the large-scale pattern is similar to the Q map at $z=0.01$.


The third step is 
the reconstruction of the initial fluctuations by the fitting 
of the polarization, $Q$ and $U$, produced by the galaxy clusters 
and the CMB temperature and polarization anisotropy, $a\lm^T$, $a\lm^E$, directly observed at the origin.
We estimate the initial fluctuations
to minimize the function given by 
\bee
f\tot = f\pol+f_{T}+f_{E}+f\pri
\ene{}
Each term in the right-hand side of the equation 
represents
the chi-square minimizations 
for fitting the polarization of the galaxy cluster $Q(x_i)$ and $U(x_i)$, 
the temperature anisotropy of the CMB $a\lm^T$ 
and the polarization anisotropy of the CMB $a\lm^E$, and the prior, respectively.

The chi-square minimizations for the polarization of the galaxy cluster $Q(x_i)$ and $U(x_i)$
can be written as
\bee
f\pol &=& \sum_{i=1}^{N\clu} \f{(Q({\vx_i})-Q({\vx_i}){}\fid)^2}{\s\pol^2}\nonumber \\
&&+\f{(U({\vx_i})-U({\vx_i}){}\fid)^2}{\s\pol^2},
\ene{}
where $Q({\vx_i})$ and $U({\vx_i})$ is the polarization produced in galaxy clusters at $\vx_i$
with the estimated initial condition 
and $Q({\vx_i}){}\fid$ and $U({\vx_i}){}\fid$ is the polarization obtained in the simulation
with adding the Gaussian noise with the variance $\s\pol$ due to
the uncertainty in the observation of $Q$ and $U$ from galaxy clusters.

We use CMB temperature anisotropy from $l=3$ to $9$ for fitting
\bee
f_{T} = \sum_{l=3}\sum_{m=-l}^l \f{(a\lm^T-a^T\lm{}\fid)^2}{\s_{T}^2},
\ene{}
where 
$a\lm^T$ is the temperature anisotropy evaluated from the estimated initial fluctuations,
$a^T\lm{}\fid$ is the one obtained from the simulation,
and $\s_{T}$ is the uncertainty in observing CMB temperature anisotropy.

For CMB polarization E-mode, $l=2$ mode is also added to the fitting function
\bee
f_{E} = \sum_{l=2}\sum_{m=-l}^l \f{(a\lm^E-a^E\lm{}\fid)^2}{\s_{E}^2}.
\ene{}
where $a\lm^E$ is the E-mode polarization anisotropy evaluated from the estimated initial fluctuations,
$a^E\lm{}\fid$ is the one obtained from the simulation,
$\s_{E}$ is the uncertainty in the observation of CMB E-mode polarization anisotropy.

To improve the accuracy of the reconstruction, we also adopt a Gaussian prior based on power spectrum $P_\phi(k)$.
\bee
f\pri = \sum^{n_k}_j\f{\ri^2(k_j)}{2P(k_j)}.
\ene{}
where $\ri^2(k)$ is the Fourier component of the estimated initial fluctuations.

Tuning the estimated initial fluctuations, $\ri^2(k)$, 
we search the set of $\ri^2(k)$ which can minimize the function $f$.
The obtained set of $\ri^2(k)$ is the Fourier component of the estimated initial fluctuations
which fit the polarization of the galaxy cluster and 
the CMB temperature and polarization anisotropy
to the values in the fiucial mock simulation.

In this process, the transfer functions are used to calculate the observable from the initial fluctuations $\phii(k_i)$. Since the transfer function depends on the cosmological parameters, different cosmologies lead to different estimates of the initial fluctuations.
In this work, we estimate the initial fluctuations with several dark energy state parameters $w=-1,\ -0.99$, and $-0.95$ in order to verify the statistical power for the dark energy state parameter
although the dark energy state parameter is fixed to $w=-1$ in the simulation.

In the last step, we calculate the $l=2$ mode temperature anisotropy $a_{2m}^T{}^{\rm{est}}(0)$ observed at the origin using the estimated initial fluctuations and compare it to the true value $a_{2m}^T{}^{\rm{true}}(0) \equiv a_{2m}^T{}\fid(0)$ calculated from the mock simulation. Note that, in the fitting process, 
we do not use the $l=2$ mode temperature anisotropy and reserve it 
for the comparison between one form the estimated initial fluctuations 
and the mock simulation data.

Up to this point, the method has been applied to a single mock simulation.
The sequence of steps is repeated one hundred times from the generation of the initial fluctuations and makes one hundred pairs of $a_{2m}^T{}^{\rm{true}}(0)$ and $a_{2m}^T{}^{\rm{est}}(0)$.

The generated $a_{2m}^T{}^{\rm{true}}(0)$ and $a_{2m}^T{}^{\rm{est}}(0)$ pairs should agree within statistical error if they are generated using the same transfer function. In application to actual observations, the cosmological parameters of the transfer function used in the estimation process should match those of the actual universe.
Thus, the larger the difference between pairs generated using different transfer functions, the more effective the method is able to constrain the cosmological parameters.

The accuracy of this method depends on errors in polarization measurements, the number of galaxy clusters, the optical depth of the clusters, and the redshift errors of the clusters. In this study, we assume the most ideal conditions, where the polarization measurement error and optical depth of the clusters are uniform $\s\pol /\e=10^{-2}\ \uk$, and the redshift error is negligible. The number of clusters used is assumed to be 6000 and randomly distributed.
The error for the CMB all-sky observation is also used as $\s_T=\s_E=10^{-2}\ \uk$.
The methodological, statistical uncertainty in this method is a complex mixture of these factors
and can be calculated from the reconstruction error in the pair 
when the correct transfer function including w=-1 is used in the estimation.
\bee
\sm^2=\f 1 N\sum^N_{i=1}\f 1 5\KAkko{|\d a_{20\ i}^T|^2+2|\d a_{21\ i}^T|^2+2|\d a_{22\ i}^T|^2}
\ene{i6}
where $N$ refers to the number of simulations used, and each $\d a_{2m}$ are difference of pairs 
\bee
\d a_{2m} = a_{2m}^T{}^{\rm{true}}(w=-1)-a_{2m}^T{}^{\rm{est}}(w=-1).
\ene{}
In Eq.~\eqref{i6}, while the $m=0$ component is a real number, the $m=1,2$ components are complex numbers, so the independent components are doubled, requiring a factor of 2 on the right side.

In the setting of our simulation with $N\clu=6000,\ {\s\pol/\tau=10^{-2}}\ \uk, \ N\side=8$ and $n_{k\rm{mode}}=60$, the methodological statistical uncertainty is
\bee
\s_\met\simeq 4.0\t 10^{-8}.
\ene{i7}
We find out that,  
even when not including all-sky CMB observations of temperature fluctuations and polarization,
almost the same values were obtained as the methodological statistical uncertainty.
Therefore, we can conclude that 
the dominant uncertainty of this reconstruction comes from the KL method.

To examine statistical power, we define the chi-square statistic for the quadrupole as
\bee
\x (w) =\f 1 {\sm^2}\kakko{|\d a_{20}^T|^2+2|\d a_{21}^T|^2+2|\d a_{22}^T|^2}.
\ene{i9}

The chi-square is an indicator to show
the goodness of fit between the cosmological model in the mock simulation and the one used for estimation.
In our case, if the equation of state of dark energy, $w$, in the estimation is identical 
to the one in the simulation, it ideally follows the chi-square distribution with a degree of freedom of five.
The chi-square values are larger when different $w$ is used in the estimation process. 

In other words, 
the cosmological parameters can be varied and the cosmology can be restricted by comparing the differences in the chi-square values $\d\x (w)=\x(w)-\x(w=-1)$.
In other words, through the comparison of the difference in the chi-square values, $\d\x (w)=\x(w)-\x(w=-1)$,
with changing $w$ in the estimation, 
we can provide the observation constraint on $w$.

\sec{Result}
In the previous study, only the polarization of the galaxy clusters was used in the fitting process to reconstruct the initial fluctuations.
In this study, we investigate the improvement in statistical power for the dark energy equation of state parameter by adding temperature anisotropy and polarization in the all-sky CMB observations.

\begin{figure}[htbp]
\centering
\includegraphics[width=1.\hsize]{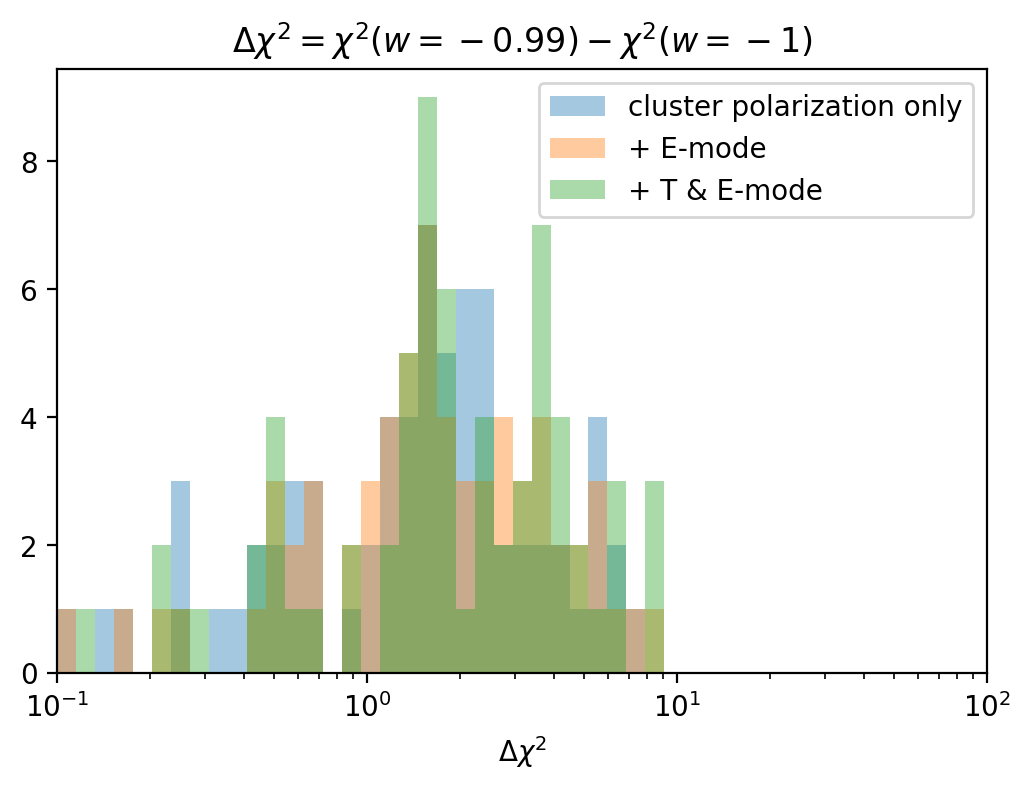}
\caption{Distribution of the difference of the chi-square statistic from the 100 simulations for $w = 0.99$. Different histograms show the cases obtained from fitting only to the polarization of galaxy clusters, fitting with the E-mode, and fitting with the E-mode and temperature anisotropies of all-sky CMB observations, as indicated in the figure.}
\label{main_w99} 
\end{figure}

We set the true equation-of-state parameters of dark energy $w=-1$. The difference in chi-square values for $w = -0.99$ is $\Kan{\x(w=-0.99)}=1.14, 1.16$, and $1.33$ respectively only galaxy clusters polarization case, 
the case with adding E-mode polarization, and the case with adding E-mode polarization and temperature anisotropy. 
We summarize the results in Table~\ref{xw99}.
Fig.\ref{main_w99} shows the histograms of $\d\x$ with 100 realizations in each case.

\begin{figure}[htbp]
\centering
\includegraphics[width=1.\hsize]{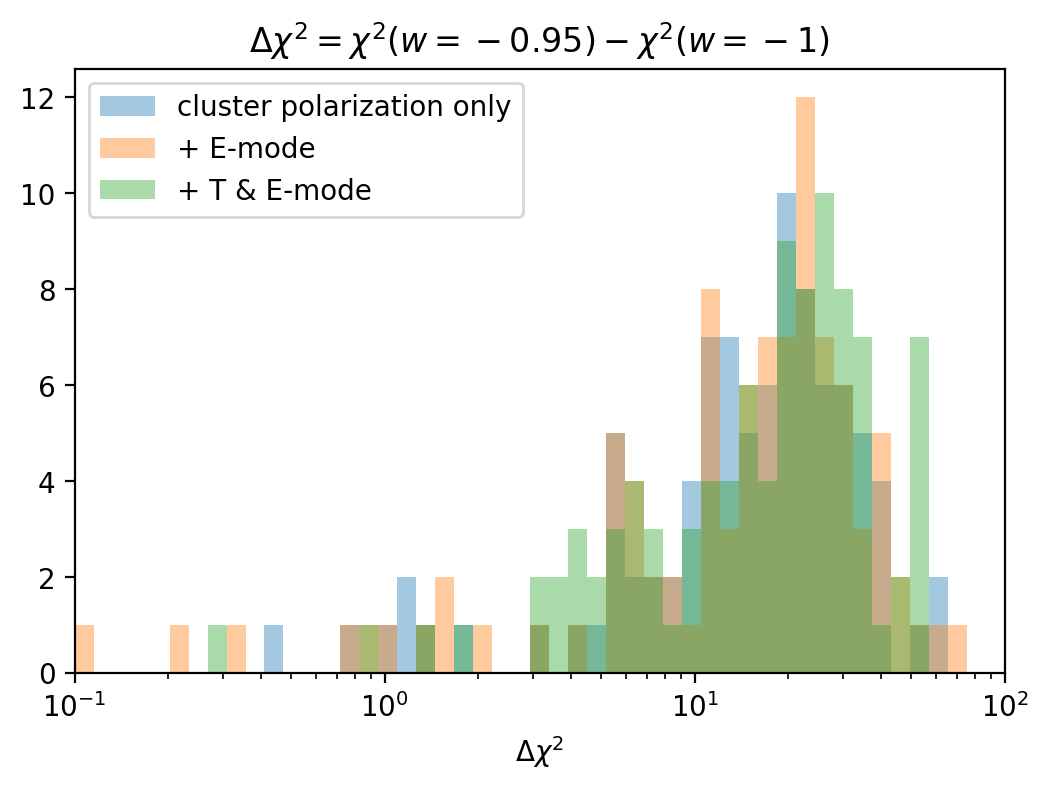}
\caption{Same as Fig.~3, but for $w = 0.95$.
}
\label{main_w95} 
\end{figure}

Also, the difference in chi-square values for $w = -0.95$ is $\Kan{\x(w=-0.95)}=16.90, 17.85$, and $19.93$  for only galaxy clusters polarization case, 
the case with adding E-mode polarization, and the case with adding E-mode polarization and temperature anisotropy, respectively. 
We summarize the results in Table~\ref{xw95}.
Fig.\ref{main_w95} shows the histograms of $\d\x$ with 100 realizations in each case.
\begin{table}[ht]
  \centering
  \begin{tabular}{lcc}
  \hline
  Observable&$\s_\met$&$\d \x$\\
  \hline \hline
   Only cluster polarization &$4.060 \t 10^{-8}$& 1.137 \\
   Cluster polarization + E-mode&$4.039 \t 10^{-8}$& 1.163 \\
   Cluster polarization + E\&T-mode&$4.014 \t 10^{-8}$&1.327 \\
   \hline
  \end{tabular}
  \caption{ $\d\x$ for parameters with $w = -0.99$}
  \label{xw99}
\end{table}

\begin{table}[ht]
  \centering
  \begin{tabular}{lcc}
  \hline
  Observable&$\s_\met$&$\d \x$\\
  \hline \hline
   Only cluster polarization &$4.060\t 10^{-8}$& 16.90 \\
   Cluster polarization + E-mode&$4.039\t 10^{-8}$& 17.85 \\
   Cluster polarization + E\&T-mode&$4.014\t 10^{-8}$&19.93\\
   \hline
  \end{tabular}
  \caption{$\d\x$ for parameters with $w = -0.95$}
  \label{xw95}
\end{table}

For both dark energy equation of state parameters, we obtained larger chi-square values when adding E-mode polarization and temperature anisotropy.

This is due to the fact that E-mode polarization and temperature anisotropy in all-sky observations are associated with the polarization produced by galaxy clusters.

Thus, combining all-sky CMB observations with the remote quadrupole technique using the polarization of galaxy clusters can more strongly constrain the cosmology.

\section{summary and discussion}
In this paper, we study how to constrain the nature of the dark energy
using the ISW effect by combining information about the CMB quadrupole
at high redshift obtained from the polarization of CMB photons passing
through a galaxy cluster based on the KL method with information about
temperature and E-mode polarization fluctuations on large angular scales
at $z = 0$. In conventional analyses based on power spectra, the SW
contribution, which is unrelated to the dark energy effect, acts like
Gaussian noise and prevents the statistical detection of the ISW effect \cite{2004PhRvD..69b7301C}. In contrast, our method can estimate and subtract the SW contribution by reconstructing the primordial density fluctuations in three dimensions. Thus, we can estimate the pure ISW effect due to dark energy.

In our previous paper, to limit the equation of state for dark energy, we used only the $z=0$ quadrupole, which is expected to correlate most with the polarization of CMB photons scattered by clusters of galaxies. However, the polarization of CMB photons scattered by clusters of galaxies, especially at high redshifts, should correlate not only with the quadrupoles but also with higher multipoles at $z=0$. Indeed, as shown in~\cite{2017PhRvD..96l3509L}, CMB polarization generated due to a galaxy cluster at a higher redshift correlates not only with the quadrupoles but also with higher multipoles of the current CMB temperature fluctuations.

Compared with the cluster polarization-only constraint, our results showed that including E-mode polarization ($l>2$) and temperature anisotropies ($l>3$) improves the constraining power for the dark energy parameter $w$ by $18$ percent if we compare $w=-1$ and $w=-0.95$ dark energy models,  assuming $6000$ clusters and polarization sensitivity of $\sigma_{pol}/\tau = 10^{-2}$. In our setup, this improvement comes almost equally from the E-mode polarization ($l>2$) and temperature anisotropies ($l>3$). The improvement is due to the fact that the information on E-mode polarization and temperature anisotropy at $z=0$ allowed us to solve part of the degeneracy between the 3D density fluctuation Fourier modes inferred from the polarization produced in galaxy clusters.

\begin{acknowledgments}
This work is supported in part by the JSPS grant numbers 18K03616,21H04467
 and JST AIP Acceleration Research Grant JP20317829 and JST
FOREST Program JPMJFR20352935 (K.I.),  
JP21K03533, and JP21H05459 (H.T.), and
JST SPRING, grant number JPMJSP2125 (H.K.).

\end{acknowledgments}



\bibliographystyle{apsrev4-2}
\bibliography{draft}

\end{document}